\documentclass[12 pt]{article}
\usepackage[dvips]{graphicx}
\usepackage{lscape}

\begin{document}
  \title{First-order Reversal Curve Analysis of Phase Transitions in
    Electrochemical Adsorption: A New Experimental Technique Suggested
    by Computer Simulations}
  \author{
    \centerline{ I. Abou Hamad$^{1,2}$, D.T. Robb$^{2}$,
      P.A. Rikvold$^{1,2,3}$}\\ 
    \centerline{\scriptsize\it $^{1}$Center for Materials Research and
    Technology and Department
    of Physics, Florida State University, Tallahassee, FL 32306-4350, USA}\\
    \centerline{\scriptsize\it $^{2}$School of Computational Science,
      Florida State University, Tallahassee, FL 32306-4120, USA}\\
    \centerline{\scriptsize\it $^{3}$National High Magnetic Field
    Laboratory, Tallahassee, FL 32310}\\
  }
  \maketitle
  \begin{abstract}
    The first-order reversal curve (FORC) method for analysis of systems
    undergoing hysteresis is applied to dynamical models of electrochemical
    adsorption. In this setting, the method can not only differentiate
    between discontinuous and continuous phase transitions, but can also
    quite accurately recover equilibrium behavior from dynamic analysis 
    for systems with a
    continuous phase transition. Discontinuous and continuous phase
    transitions in a two-dimensional lattice-gas model are compared using
    the FORC method. The FORC diagram for a discontinuous
    phase transition is characterized by a negative (unstable)
    region separating two positive (stable) regions, while such a negative
    region does not exist for continuous phase transitions.
    Experimental data for FORC analysis could easily be obtained
    by simple reprogramming of a potentiostat designed for cyclic-voltammetry
    experiments.
    
  \end{abstract}
  
      
      \section{Introduction}
      \label{sec:I}
      Recent technological developments in electrochemical deposition
      have made possible experimental studies of atomic-scale
      dynamics~\cite{Tansel:06}. It is therefore now both timely and
      important to develop new computational methods for the analysis
      of experimental adsorption dynamics. In this paper we apply one
      such analysis technique, the first-order reversal curve (FORC) method,
      to analyze model systems with continuous and discontinuous phase
      transitions. We propose that the method can be a useful
      new experimental tool in surface electrochemistry.
      
      The FORC method was originally conceived~\cite{kn:mayergoyz86} in
      connection with the Preisach model of magnetic hysteresis. It
      has since been applied to a variety of magnetic systems, ranging
      from magnetic recording media and nanostructures to
      geomagnetic compounds, undergoing \textit{rate-independent} (i.e.,
      very slow) magnetization reversal~\cite{kn:pike99}. Recently, there have
      also been several FORC studies of \textit{rate-dependent}
      reversal~\cite{kn:enachescu05,kn:fecioru-morariu04,kn:robb05}.
      Here we introduce and apply the FORC method in an electrochemical
      context. For completeness, a brief translation to magnetic language
      is found in the Appendix.
      
      We apply FORC analysis to rate-dependent adsorption in two-dimensional
      lattice-gas models of electrochemical deposition. Specifically,
      we study a lattice-gas model with attractive nearest-neighbor
      interactions (a simple model of underpotential deposition, UPD),
      being driven across its discontinuous phase transition by a
      time-varying electrochemical potential. In addition, we consider
      a lattice-gas model with repulsive lateral interactions and
      nearest-neighbor exclusion (similar to the model of halide adsorption on
      Ag(100), described in
      Refs.~\cite{AbouHamad:04,MitchellSS:01,AbouHamad:03,AbouHamad:05}), being
      similarly driven across its continuous phase transition.
      
      The rest of this paper is organized as follows. In Sec.~\ref{sec:forc}
      the FORC method is explained. The model used for
      both systems with continuous and discontinuous transitions is briefly
      discussed in Sec.~\ref{sec:M}. In Sec.~\ref{sec:S} the dynamics of
      systems with a discontinuous phase transition are studied using
      Kinetic Monte Carlo (KMC) simulations, as well as a mean-field model. The
      dynamics of systems with a continuous phase transition are studied
      in Sec.~\ref{sec:C}. Finally, a comparison between
      the two kinds of phase transitions and our conclusions are presented
      in Sec.~\ref{sec:conc}.
      
      \section{The FORC Method} 
      \label{sec:forc}
\begin{figure}
	\begin{center}
	  \includegraphics[width=.4\textwidth]{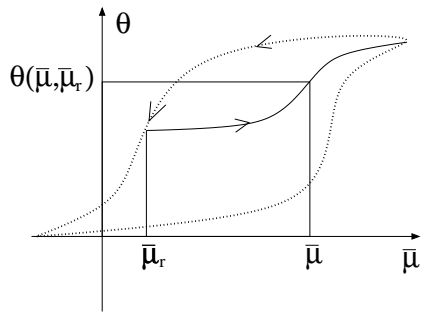}
	  \caption[Schematic diagram of a FORC curve]{Schematic diagram of
            a FORC curve.}
	  \label{fig:loop}
	\end{center}
      \end{figure}

      For an electrochemical adsorption system, the FORC method consists of
      saturating the adsorbate coverage $\theta$ in a strong positive
      (for anions; negative for cations)
      electrochemical potential (proportional to the electrode potential)
      and, in each case starting from
      saturation, decreasing the potential to a series of progressively
      more negative ``reversal potentials'' $\bar{\mu}_r$ (Fig.~\ref{fig:loop}).
      Subsequently, the potential is increased back to the saturating
      value~\cite{kn:pike99}. It is thus a simple generalization of the
      standard cyclic voltammetry (CV) method, in which the negative return
      potential is decreased for each cycle. This produces a family of
      FORCs, $\theta (\bar{\mu}_r, \bar{\mu})$, where $\theta$ is the
      adsorbate coverage, and where $\bar{\mu }$ is the instantaneous
      potential during the increase back toward saturation.
      Although we shall not discuss this further here, it is of course also
      possible to fix the negative limiting electrode potential and change
      the positive return potential from cycle to cycle.
      
      It is further useful to calculate the FORC distribution,
      \begin{equation}
	\rho=- \frac{1}{2} \frac{\partial^2 \theta}{\partial\bar{\mu}_r\,
	  \partial\bar{\mu}}\,, 
	\label{forc.definition}
      \end{equation}
      which measures the sensitivity of the dynamics to the progress
	of reversal along the major loop.\footnote{Note that to
	normalize the FORC distribution, the term $\frac{1}{2} \delta
	(\bar{\mu}~-~\bar{\mu}_r ) \frac{\partial \theta
	(\bar{\mu}_r,\bar{\mu})}{\partial \bar{\mu}} |_{\bar{\mu} \to
	  {\bar{\mu}_r}^+}$ must be added to
	Eq.~(\ref{forc.definition})~\cite{kn:pike03}. Here we consider
	the distribution only away from the line $\bar{\mu} =
	\bar{\mu}_r$. The 
	additional term could be found from the major loop.}
	The FORC distribution is usually displayed as a contour plot called a
	`FORC diagram.' A positive value of $\rho$ indicates that the
	corresponding reversal curves are converging with increasing
	$\bar \mu$, while a negative value indicates divergence. Some
	preliminary results of this work have been submitted for
	publication elsewhere~\cite{AbouHamad:06}.
	
	\section{Model}
	\label{sec:M}
	KMC simulations of lattice-gas models, where a
	Monte Carlo (MC) step corresponds to an attempt to cross a
	free-energy barrier, have been used to simulate the kinetics of
	electrochemical systems with
        discontinuous~\cite{AbouHamad:03,AbouHamad:05,Frank:05,Frank:06} or
	continuous~\cite{AbouHamad:04,Mitchell:02} phase transitions in two
	dimensions. The energy associated with a lattice-gas configuration is
	described by the grand-canonical effective Hamiltonian for an
	$L\times L$ square system of adsorption sites,
	\begin{equation}
	  {\mathcal{H}} = - \sum_{i<j} \phi_{ij} c_{i} c_{j} - \bar{\mu} 
	  \sum_{i=1}^{L^2}c_{i}\; ,
	  \label{eq:H-II}
	\end{equation}
	where $\sum_{i<j}$ is a sum over all pairs of sites, $ \phi_{ij} $ are
	the lateral interaction energies between particles on the $i$th and
	$j$th sites measured in meV/pair, and $ \bar{ \mu } $ is the
	electrochemical potential measured in meV/atom. The local
	occupation variables $ c_{i} $ can take the values 1 or 0, depending
	on whether site $ i $ is occupied by an ion (1) or solvated (0).
	The sign convention is chosen such that $\bar\mu > 0$ favors
	adsorption, and negative values of $\phi_{ij}$ denote repulsion
	while positive values denote attraction between adsorbate particles
	on the surface. In addition
	to adsorption/desorption steps, we include diffusion steps
	with a comparable free-energy barrier~\cite{AbouHamad:04}. 
	
	In each time step of the KMC simulation, an adsorption site is
	chosen at random and
	the transition rates from the present configuration to a set of new
	configurations (desorption, diffusion) are calculated. A weighted
	list for accepting each of these moves is constructed using
	Eq.~(\ref{eq:P}) below, to calculate the probabilities
	$R\rm{(F|I)}$ of the individual moves between the initial state $\rm I$
	and final state $\rm F$. The probability for
	the system to stay in the initial configuration is consequently
	$R\rm{(I|I)}=1-$$\Sigma_{\rm F\neq
	I}R\rm{(F|I)}$~\cite{AbouHamad:04,MitchellSS:01}.
	
	Using a thermally activated, stochastic
	barrier-hopping picture, the energy of the transition state for a
	microscopic change from an initial state $\rm I$ to a final state
	$\rm F$ is approximated by the symmetric Butler-Volmer
	formula~\cite{Brown:99,Kang:89,Buendia:04}
	\begin{equation}
	  U_{\rm T_{\lambda}}=\frac{U_{\rm I}+U_{\rm
	F}}{2}+\Delta_{\lambda} \; .
	\end{equation}
	Here $U_{\rm I}$ and $U_{\rm F}$ are the energies of the
	initial and final states, respectively, $\rm T_{\lambda}$ is
	the transition state for process $\lambda$, and
	$\Delta_{\lambda}$ is a ``bare'' barrier associated with
	process $\lambda$. This process can here be either
	nearest-neighbor diffusion ($\Delta_{\rm nn}$),
	next-nearest-neighbor diffusion ($\Delta_{\rm nnn}$), or
	adsorption/desorption ($\Delta_{\rm a/d}$). The probability
	for a particle to make a transition from state $\rm I$ to
	state $\rm F$ is approximated by the one-step Arrhenius
	rate~\cite{Brown:99,Kang:89,Buendia:04}
	\begin{equation}
	  \mathcal{R}({\rm F}|{\rm I})= \nu \exp\left(-\frac{(U_{\rm T_{\lambda}}-U_{\rm I})}{k_{\rm B}T}\right) =  \nu \exp
	  \left(-\frac{\Delta_{\lambda}}{k_{\rm B}T}\right)
	  \exp\left(-\frac{U_{\rm F}-U_{\rm I}}{2k_{\rm
	  B}T}\right),\label{eq:P}
	\end{equation}
	where $\nu$ is the attempt frequency, which sets the overall
	timescale for the simulation. The electrochemical potential $\bar \mu$,
	which is proportional to the electrode potential, is increased
	monotonically, preventing the system from reaching equilibrium at the
	instantaneous value of $\bar \mu$.
	
	Independent of the diffusional degree of freedom, attractive
	interactions ($\phi_{ij}>0$) produce a discontinuous phase transition
	between a low-coverage phase at low $\bar\mu$, and a high-coverage
	phase at high $\bar\mu$. In contrast, repulsive interactions
	($\phi_{ij}<0$) produce a continuous 
	phase transition between a low-coverage disordered phase
	for low $\bar\mu$, and a high-coverage, ordered phase for high
	$\bar\mu$. Examples of systems with a discontinuous phase
	transition include underpotential
	deposition~\cite{Frank:05,Frank:06,BrownJES:99}, while the
	adsorption of halides on
	Ag(100)~\cite{AbouHamad:04,MitchellSS:01,Mitchell:02,Ocko:97,Wandlowski:01}
	are examples of systems with a continuous phase transition. 
	\section{Discontinuous Phase Transition}
	\label{sec:S}
\begin{figure}
	  \begin{center}
	    \includegraphics[width=.6\textwidth]{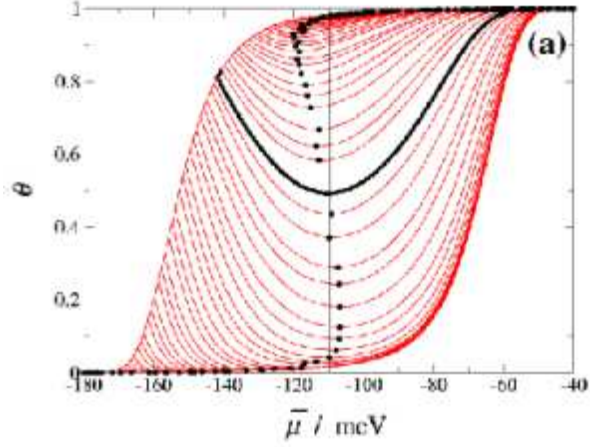}
	    \mbox{}
	    \vspace{0.4in}
	    \mbox{}
	    \includegraphics[width=.6\textwidth]{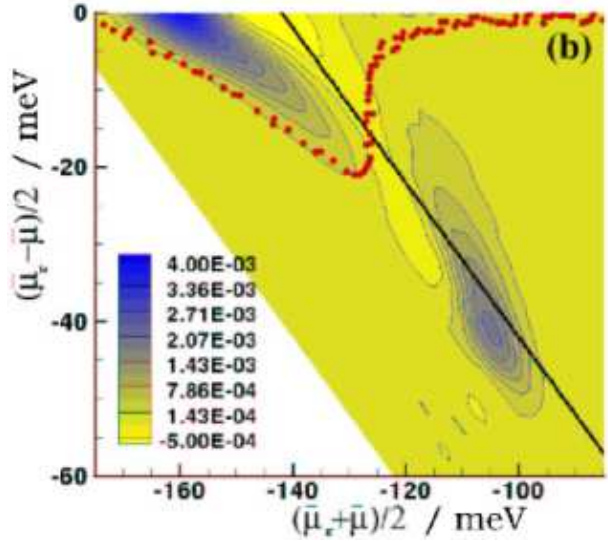}
	  \end{center}
	  \caption[({\bf a}) First-order reversal curves (FORCs) and FORC
	    distribution for a discontinuous phase transition]{ (Color online.)
	    First-order reversal curves (FORCs) for a discontinuous
	    phase transition. The vertical line shows the position of
	    the coexistence value, $\bar{\mu} = \bar{\mu}_0$. The
	    minima  of each FORC are also shown (circles).
	    ({\bf b})
	    FORC diagram generated from the family of FORCs shown in
	    ({\bf a}). The positions of the FORC minima are also shown
	    (circles). The straight line corresponds to the FORC for
	    which the minimum lies at the coexistence value $\bar{\mu}
	    = \bar{\mu}_0$ (thick curve in ({\bf a})). After
	    Ref.~\cite{AbouHamad:06}.
	  }
	  \label{fig:fig1}
	\end{figure}

	A two-dimensional lattice gas with attractive adsorbate-adsorbate
	lateral interactions that cause a discontinuous phase transition
	is a simple model of electrochemical underpotential
	deposition~\cite{Frank:05,Frank:06,BrownJES:99,Ramos:99}. 
	Using a lattice-gas model with attractive interactions on an
        $L \times L$ lattice with $L=128$, a
	family of FORCs were simulated, averaging over ten
	realizations for each reversal curve at room temperature.
	The lateral interaction
	energy (restricted to nearest-neighbor)
	was taken to be $\phi_{ij}= \phi_{\rm nn} = 55 $\,meV, where
	the positive value indicates nearest-neighbor attraction. For
	this value of $\phi_{\rm nn}$, room temperature
	corresponds to $T=0.8 T_c$, where $T_c$ is the critical
	temperature. The barriers for adsorption/desorption and
	diffusion (nearest-neighbor only) were $\Delta_{\rm
	a/d}=\Delta_{\rm nn} = 150$\,meV, corresponding to relatively slow
        diffusion~\cite{Frank:06}. Simulation runs with faster
	diffusion ($\Delta_{\rm nn} = 125$\,meV) and the same
	adsorption/desorption barrier showed little difference
	from Fig.~\ref{fig:fig1}, indicating that diffusion effects are
        not significant for this model.
	The reversal electrochemical potentials $\bar{\mu}_r$
	associated with the reversal curves were separated by $1$\,meV
	increments in the interval $[-200\,{\rm meV},0\,{\rm meV}]$, and the
	field-sweep rate was constant at $|{\rm d}\bar{\mu}/{{\rm d}t}| = 0.3 
	\mathrm{meV}/\mathrm{MCSS}$. The FORCs are shown in
	Fig.~\ref{fig:fig1}({\bf a}), with a vertical
	line indicating the position of the coexistence value of the
	electrochemical potential, $\bar{\mu}_0=-110$\,meV, and circles
	showing the position of the minimum of each FORC.
	
	In a simple Avrami's-law analysis, the FORC minima all lie at
	$\bar{\mu}=\bar{\mu}_0$~\cite{kn:robb05}. However, in the
	simulations the minima are displaced. For $\theta>0.5$, the minima
	occur at $\bar{\mu}<\bar{\mu}_0$, precisely at the
	points where the tendency to phase order, which drives local regions
	of the system toward the nearby metastable state ($\theta\approx1$),
	is momentarily balanced by the electrochemical potential, which
	drives the system toward the distant stable state
	($\theta\approx0$). For $\theta<0.5$, the stable and
	metastable states are $\theta\approx1$ and $\theta\approx0$,
	respectively, and the same balancing effect explains the FORC minima
	occurring at $\bar{\mu}>\bar{\mu}_0$.
	\begin{figure}
	  \begin{center}
	    \includegraphics[width=.65\textwidth]{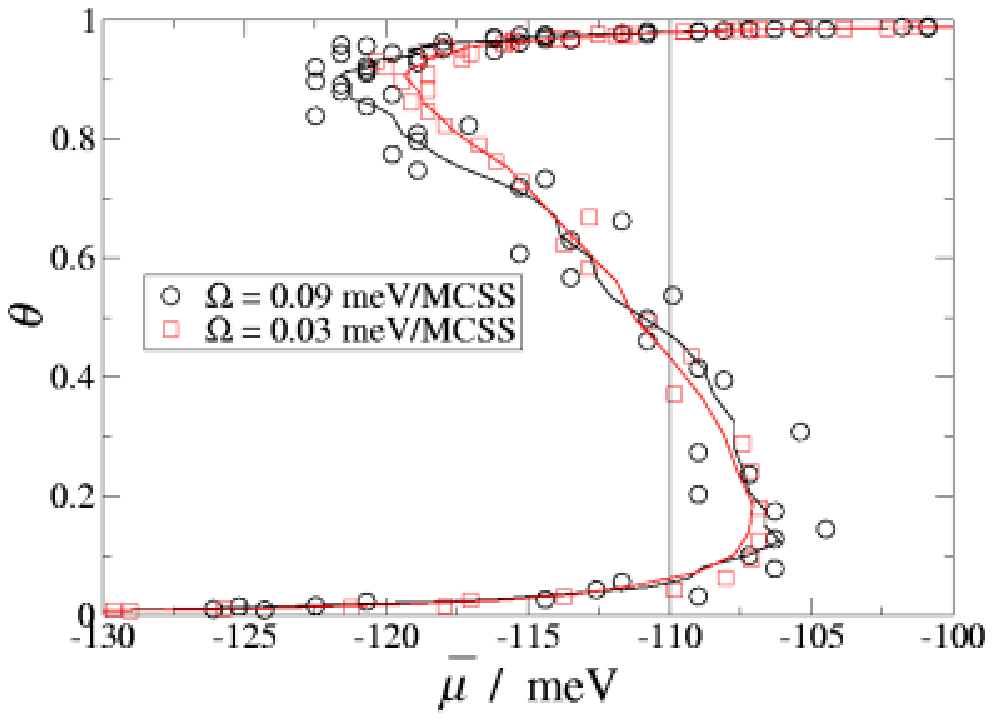}
	    \caption[Dependence of the FORC minima on the sweep
	      rate]{FORC minima dependence on sweep rate. The figure shows FORC
	      minima for two families of FORCs with different sweep rates
	      ($\Omega=0.03$ and $0.09$\,meV/MCSS). The
	      lines are guides to the eye, obtained by smoothing the
	      data using a Savitzky-Golay
	      method~\cite{Savitzky:64,Press:nr} with a first-order
	      polynomial and a window of 5 points.
	    }
	    \label{fig:vanloop}
	  \end{center}
	\end{figure}
	
	The net effect
	is a `back-bending' of the curve of minima, as seen in
	Fig.~\ref{fig:fig1}({\bf a}).
\begin{figure}
	  \begin{center}
	    \includegraphics[width=.65\textwidth]{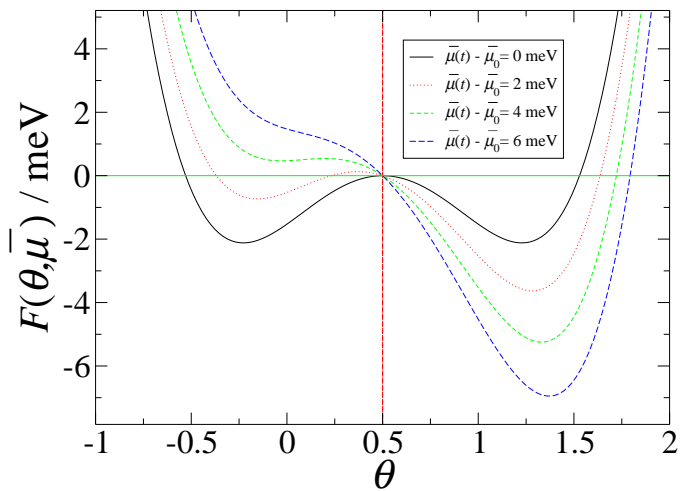}
	    \caption[Free energy of a fourth-order Ginzburg-Landau form]{
	      (Color online.) Free energy of a fourth-order
	      Ginzburg-Landau form as a function of $\theta$, given by
	      Eq.~(\ref{eq:free.energy}). The parameters were
	      calculated as $a=30.2$\,meV, $b=16.0$\,meV,
	      $\theta_c=0.5$, and $(\bar{\mu}-\bar{\mu}_0) = 0,2,4,$
	      and 6\,meV. A metastable state at $\theta<0.5$ exists
	      for the curves with $(\bar{\mu}-\bar{\mu}_0) =
	      0,2,4$\,meV, but has disappeared in the curve
	      $(\bar{\mu}-\bar{\mu}_0) = 6$\,meV. The metastable
	      minimum disappears at the so-called spinodal potential,
	      $\bar{\mu}_{\rm sp}-\bar{\mu}_0=\frac{2b}{3}
	      \sqrt{\frac{b}{3a}}\approx4.48$\,meV. 
	    }
	    \label{fig:Fvstheta}
	  \end{center}
	\end{figure}
	The definition in Eq.~(\ref{forc.definition}) implies that the FORC
	distribution $\rho$ should be negative
	in the vicinity of the back-bending. This can
	be seen in Fig~\ref{fig:fig1}({\bf b}), where the FORC
	distribution is plotted against the variables $\bar{\mu}_b =
	(\bar{\mu}_r + \bar{\mu})/2$  and $\bar{\mu}_c = (\bar{\mu}_r
	- \bar{\mu})/2$. The negative values of $\rho$ reflect a local
	divergence of the FORCs away from each other, which can be
	considered a dynamical instability, caused by the competition
	between the tendency to phase order and the effect of the
	electrochemical potential. It is important to note that, when
	the potential sweep is stopped suddenly, after the dynamical
        instability played out on short timescales, the system is
	observed to relax reliably to the stable state at that
	potential on large timescales. The only exception is the
	point ($ \bar{\mu} = \bar{\mu}_0 , \theta = 0.5$) along the FORC
	indicated by a bold line in Figs. \ref{fig:fig1}({\bf a}) and
	\ref{fig:fig1}({\bf b}). It is also interesting to note that
	the curve connecting the minima of the FORCs resembles the van
	der Waals loop in the mean-field isotherm of a fluid
	system~\cite{Castellan}, but with an asymmetrical shape about
	the point ($\bar{\mu} = \bar{\mu}_0, \theta = 0.5$) and with
	a sweep-rate dependent shape (see Fig.~\ref{fig:vanloop}).
	
	We next explore this connection with the van der Waals loop in
	numerical solutions of a kinetic mean-field model.  
	We can describe the competition between the phase-ordering 
	and the influence of the potential explicitly, using a time-dependent
	mean-field model of the dynamics. The free energy
	is approximated by a fourth-order Ginzburg-Landau
	form~\cite{Landau:00},
	\begin{equation}
	  \label{eq:free.energy}
	  F(\theta,\bar{\mu}(t)) = a\frac{(\theta-\theta_c)^4}{4} -
	  b\frac{(\theta-\theta_c)^2}{2} -
	  (\theta-\theta_c)(\bar{\mu}(t)-\bar{\mu}_0).
	\end{equation}
	A plot of Eq.~(\ref{eq:free.energy}) is shown in
	Fig.~\ref{fig:Fvstheta} for $\theta_c=0.5$, with the values of
	$a$ and $b$ calculated as described below and with
	$\bar\mu(t)-\bar{\mu}_0=0,2,4,$ and $6$\,meV. In the
	noise-free (zero-temperature) case, the dynamics are given by
	\begin{equation}
	  \frac{{\rm d}\theta}{{\rm d}t} = - \frac{1}{\gamma}
	  \frac{\partial F}{\partial \theta} = -\frac{1}{\gamma}
	  \left( a (\theta-\theta_c)^3  - b(\theta-\theta_c) -
	  (\bar{\mu}(t)-\bar{\mu}_0) \right),
	  \label{tdgl.motion}
	\end{equation}
        where $\gamma$ is a phenomenological damping parameter.
	The parameters to use in this mean-field model can
	be determined directly from the simulated family of FORCs. On
	the curve of minima of the FORCs, ${\rm d}\theta/{{\rm d}t}=0$, 
        which implies
	\begin{equation}
	  a (\theta-\theta_c)^3  - b(\theta-\theta_c) - 
	  (\bar{\mu}-\bar{\mu}_0) = 0\;. 
	  \label{line.minima.analytic}
	\end{equation}
	\begin{figure}
	  \begin{center}
	    \includegraphics[width=.6\textwidth,angle=0]{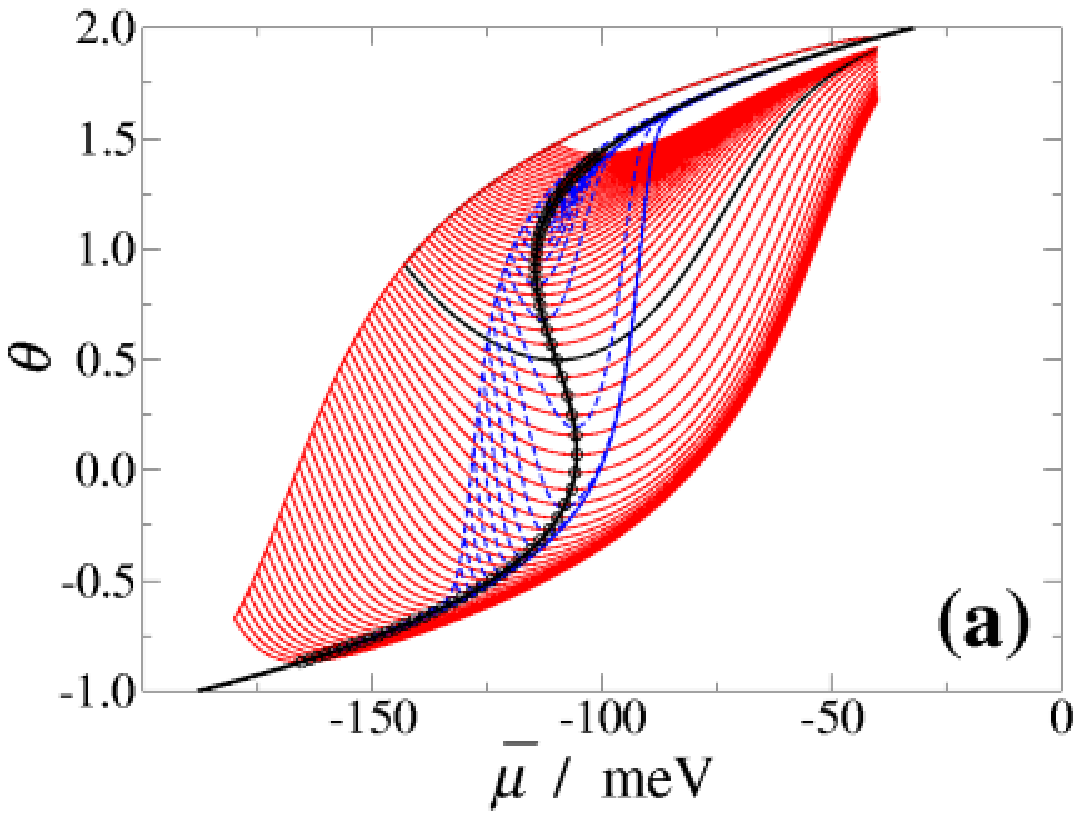}
	    \mbox{}
	    \vspace{0.05in}
	    \mbox{}
	    \includegraphics[width=.5\textwidth,angle=0]{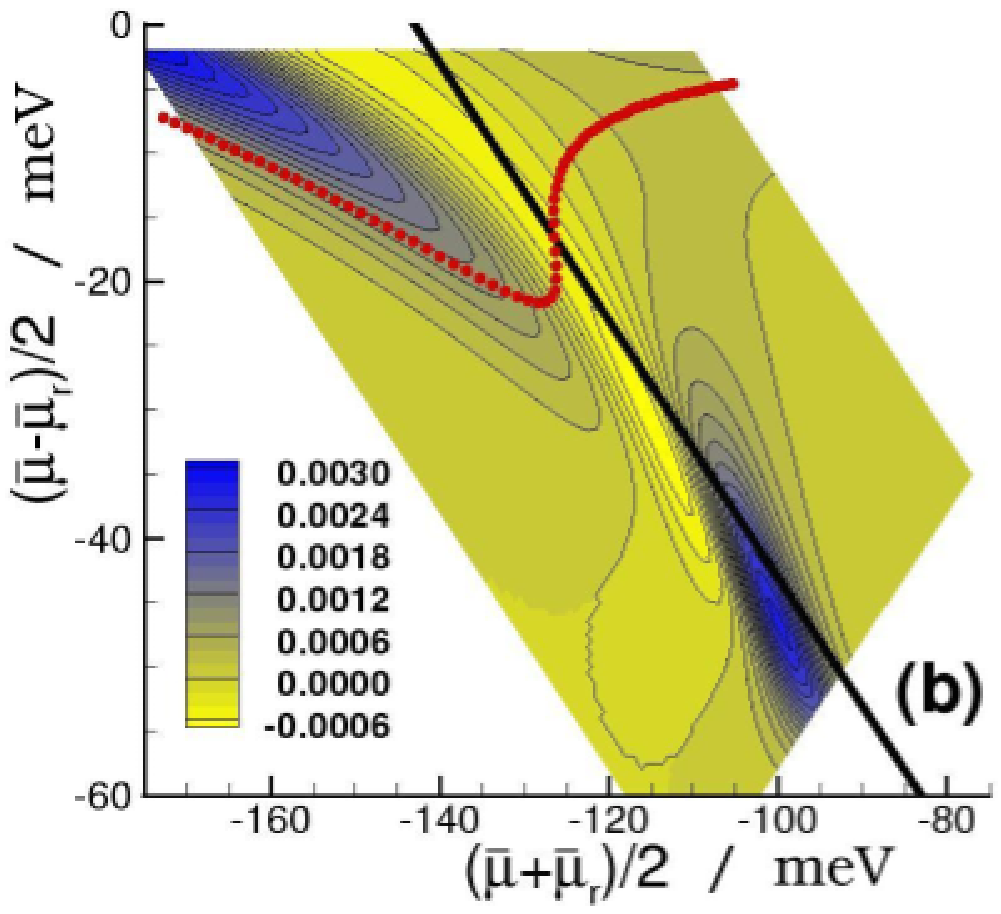}
	  \end{center}
	  \caption[Reversal behavior of a dynamic mean-field model,
	    FORCs and FORC distribution]{(Color online.) Reversal
	    behavior of a dynamic mean-field model: (a) Families of
	    FORCs. (b) FORC distribution for one family. The
	    parameters were calculated as $a=30.2$\,meV, $b=16.0$\,meV, and
	    $\gamma = 3720$\,MCSS\,meV to
	    fit the results of Fig.~\ref{fig:fig1}({\bf a}). In ({\bf a}), the
	    solid lines show the FORCs for faster sweep rate
	    ($\Omega = 0.3$\,meV/MCSS), and the dashed lines show the
	    FORCs for a slower sweep rate ($\Omega = 0.03$\,meV/MCSS). The
	    thick line is the analytical result,
	    Eq.~(\ref{line.minima.analytic}), for the line of minima,
	    while the circles show the actual minima from the
	    numerical integration. In (b), the full FORC distribution
	    is plotted vs ($\bar{\mu}_b , \bar{\mu}_c$) (see text) for
	    the FORCs with the faster sweep rate
	    ($\Omega=0.3$\,meV/MCSS). The straight line corresponds to
	    the FORC for which the minimum lies at the coexistence
	    value $\bar{\mu} = \bar{\mu}_0$ (thick curve in ({\bf a})). 
	    The appearance is similar to Fig~\ref{fig:fig1}({\bf b}).
	  } 
	  \label{tdgl.forcs}
	\end{figure}

	Differentiating with respect to $\bar{\mu}$ and solving
	for ${\rm d}\theta/{\rm d}\bar{\mu}$ gives
	\begin{equation}
	  \frac{{\rm d}\theta}{{\rm
	  d}\bar{\mu}}=\frac{1}{3a(\theta-\theta_c)^2-b} \; . 
	\end{equation}
	For $\theta=\theta_c$, this yields 
	$b = - \left({\rm d}\theta/{\rm
	d}\bar{\mu}|_{\theta=\theta_c}\right)^{-1}$. The so-called
	spinodal potential $\bar{\mu}_{\rm sp}$, and spinodal coverage
	$\theta_{\rm sp}$, occur where ${\rm d}\theta/{\rm
	d}\bar{\mu}$ diverges, so that
	\begin{equation}
	  3a(\theta_{\rm sp}-\theta_c)^2-b = 0\,,
	\end{equation}
	yielding
	\begin{equation}
	  a = b/[3(\theta_{\rm sp}-\theta_c)^2]\,.
	\end{equation}
	
	The damping parameter $\gamma $ can be related to the
	``coercive potential'' $\bar{\mu}_{\rm coer}$ (The value of the
	electrochemical potential at which $\theta=\theta_c$ on the
	major loop), the slope $\frac{{\rm d}\theta}{{\rm
	d}\bar{\mu}}|_{\bar{\mu}=\bar{\mu}_{\rm coer}}$, and the sweep
	rate ${\rm d}\bar{\mu}/{\rm d}t$ as
	\begin{equation}
	  \left. \frac{{\rm d}\theta}{{\rm d}\bar{\mu}}
	  \right|_{\bar{\mu}=\bar{\mu}_{\rm coer}} = \frac{{\rm
	      d}\theta/{\rm d}t |_{\theta=\theta_c}}{{\rm d}\bar{\mu}/{\rm
	      d}t} = \frac{(\bar{\mu}_{\rm coer}-\bar{\mu}_0)}{\gamma \;
	    {\rm d}\bar{\mu}/{\rm d}t}\,, 
	\end{equation}
	where we have used Eq.~(\ref{tdgl.motion}) in the last step.  
	
\begin{figure}
	  \begin{center}
	    \includegraphics[width=.65\textwidth]{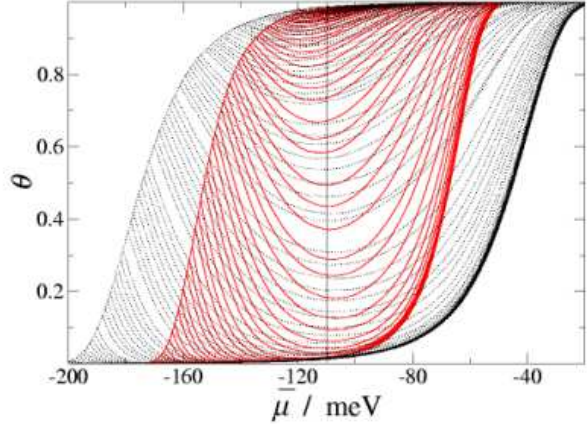}
	    \caption[KMC simulation results for FORCs
              with different sweep rates]{(Color online.)
              KMC simulation results for FORCs
              with different sweep rates, $\Omega=0.03$\,meV/MCSS (solid lines)
              and $0.09$\,meV/MCSS (dotted lines). The vertical line
              shows the coexistence value, $\bar{\mu} = \bar{\mu}_0$.
	      The coercive potential for the FORCs with
              the slower sweep rate is shifted toward
              the $\bar{\mu}=\bar{\mu}_0$ line.
	    }
	    \label{fig:sweep}
	  \end{center}
	\end{figure}

	In Fig.~\ref{tdgl.forcs}({\bf a}), we show the family of FORCs
	obtained by numerical integration of the mean-field model,
	with parameters determined by a fit to the simulated family of
	FORCs in Fig.~\ref{fig:fig1}({\bf a}). The line of minima in
	Fig.~\ref{tdgl.forcs}({\bf a}) follows the analytical curve
	defined by Eq.~(\ref{line.minima.analytic}), as
	expected. However, in contrast to Fig.~\ref{fig:fig1}({\bf
	a}), the line of minima is symmetric about the point
	$(\bar{\mu}=\bar{\mu}_0, \theta=\theta_c)$. In addition, in
	KMC simulations at a slower sweep rate, we found that both the
	coercive potential (Fig.~\ref{fig:sweep}) and the line of
	minima~(Fig.~\ref{fig:vanloop}) of the FORCs were shifted toward
	$\bar{\mu}=\bar{\mu}_0$. In contrast, in the mean-field model at 
	a slower sweep rate, the coercive field decreases, but the line of
	minima remains unchanged (as it must, since
	Eq.~(\ref{line.minima.analytic}) is independent of the sweep rate).
	Preliminary simulations indicate that including thermal noise in
        the mean-field model may further improve the agreement
	with the MC results.
	\section{Continuous Phase Transition}
	\label{sec:C}
	\begin{figure}
	  \vspace{0.4truecm}
	  \begin{center}
	    \includegraphics[width=.6\textwidth]{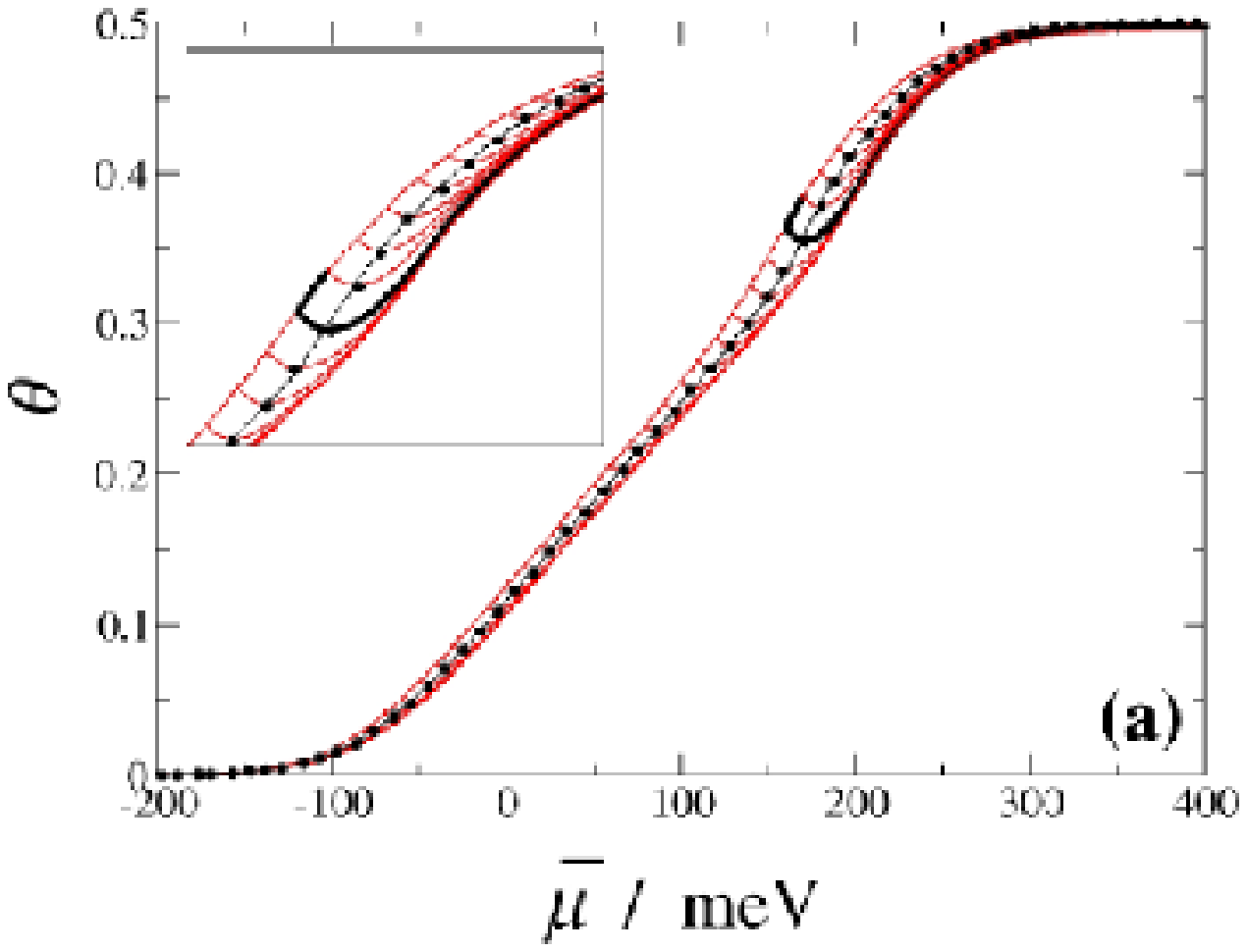}
	    \mbox{}
	    \vspace{0.1in}
	    \mbox{}
	    \includegraphics[width=.55\textwidth]{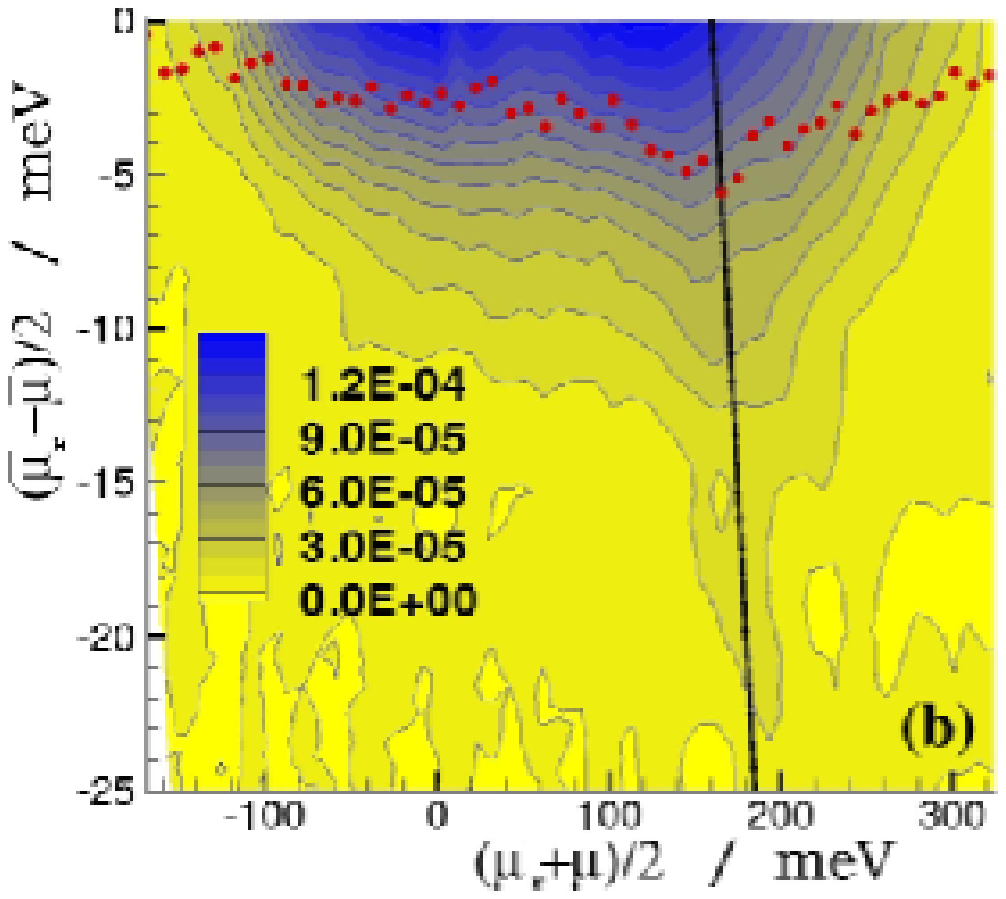}
	  \end{center}
	  \caption[First-order reversal curves (FORCs) and FORC
	    distribution for a continuous phase transition]{ (Color online.)
	    ({\bf a})
	    First-order reversal curves (FORCs) for a continuous phase
	    transition simulated at a slow scan rate $\Omega=0.0003$\,meV/MCSS.
	    The thin black middle line shows the equilibrium curve. The inset
	    is a magnification of the critical region. The minima 
	    of each FORC are also shown (black filled circles). The thick
     	    black line shows the first FORC which dips below the critical
	    coverage.
	    ({\bf b})
	    FORC diagram generated from the FORCs shown in ({\bf a}). The
	    positions of the FORC minima are also shown (circles). The
	    straight line corresponds to the first FORC for which the
	    minimum dips below the critical coverage. After
	    Ref.~\cite{AbouHamad:06}.
	  }
	  \label{fig:fig2}
	\end{figure}
	Using the same Hamiltonian, but with long-range repulsive interactions
	and nearest-neighbor exclusion as discussed in
	Ref.~\cite{AbouHamad:04}, KMC simulations were
	used to produce the family of FORCs for a continuous phase
	transition. The reversal potentials $\bar{\mu}_r$ were
	separated by $10$\,meV increments in the interval $[-200\,{\rm
	meV},400\,{\rm meV}]$. As in Ref.~\cite{AbouHamad:04}, the repulsive
	$1/r^3$ interactions, with nearest-neighbor exclusion and
	$\phi_{\rm nnn} = -21$\,meV, are calculated with exact
	contributions for $r_{ij} \le 3$, and using a mean-field
	approximation for $ r_{ij} > 3$. The barriers for
	adsorption/desorption and nearest- and next-nearest-neighbor
	diffusion, are $\Delta_{\rm a/d} = 300$\,meV, $\Delta_{\rm nn}
	= 100 $\,meV, and $\Delta_{\rm nnn} = 200$\,meV,
	respectively~\cite{AbouHamad:04}. Larger values of the
	diffusion barrier were also used to study the effect of
	diffusion on the dynamics. A continuous phase transition
	occurs between a disordered state at low coverage and an
	ordered state at high coverage~\cite{Ocko:97,Wandlowski:01}.
        The FORCs and the FORC diagram are shown in
	Fig.~\ref{fig:fig2}. Also indicated in
	Fig.~\ref{fig:fig2}({\bf a}) are the FORC minima and the
	equilibrium isotherm.
	
	Note that the FORC minima in Fig.~\ref{fig:fig2}({\bf a}) 
        lie directly on the equilibrium
	isotherm. This is because such a system has one stable state
	for any given value of the potential, as defined by the continuous
	equilibrium curve. The uniformly positive value of the FORC
	distribution in Fig.~\ref{fig:fig2}({\bf b}) reflects the convergence
	of the family of FORCs with increasing $\bar{\mu}$. This convergence
	results from relaxation toward the equilibrium isotherm, at a
	rate which increases with the distance from equilibrium. It is
	interesting to note that, while it is difficult to see at this
	slow scan rate, the rate of approach to equilibrium
	decreases greatly along the first FORC that dips below the
	critical coverage $\theta_c \approx 0.36$ (shown in bold in
	Fig.~\ref{fig:fig2}({\bf a})). The FORCs that lie completely
	in the range $\theta > 0.36$ never enter into the disordered
	phase, and thus their approach to equilibrium is not hindered
	by jamming. This is a phenomenon that occurs when further
	adsorption in a disordered adlayer is hindered by the
	nearest-neighbor exclusion. As a result, extra diffusion steps
	are needed to make room for the new adsorbates, and the system
	follows different dynamics than a system with an ordered
	adlayer~\cite{Privman:93}. The FORCs that dip below
	$\theta_c=0.36$ enter into the disordered phase, and thus
	their approach to equilibrium is delayed by jamming. This is
	reflected in the FORC diagram by the Florida-shaped
	``peninsula'' centered around this FORC in
	Fig.~\ref{fig:fig2}({\bf b}).

\begin{figure}
	  \vspace{0.4truecm}
	  \begin{center}
	    \includegraphics[width=.65\textwidth]{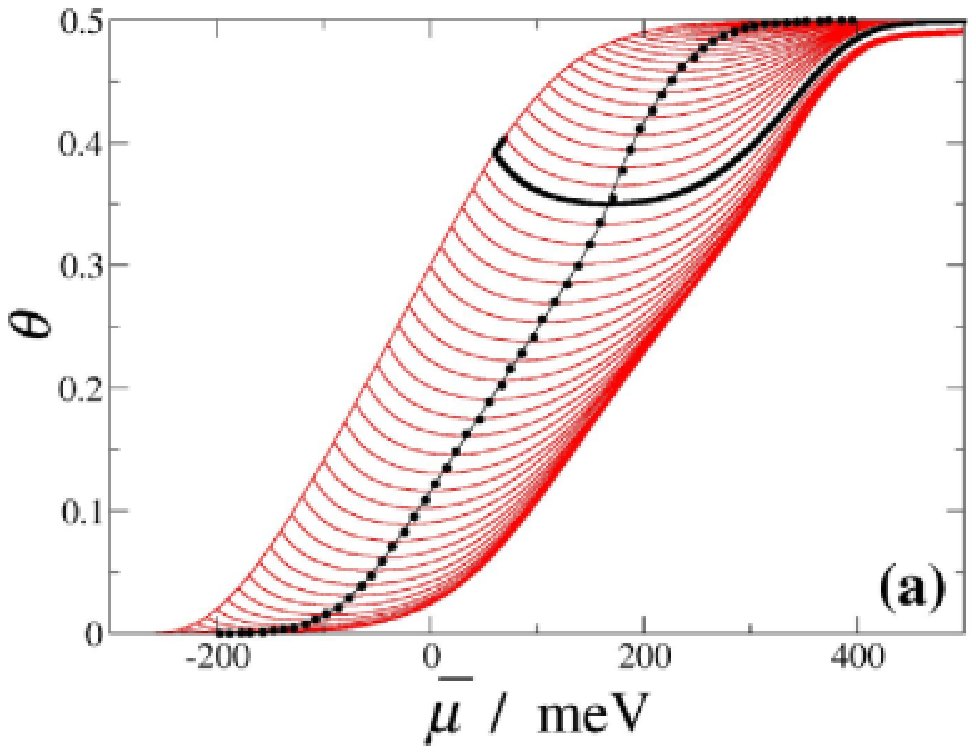}
	    \mbox{}
	    \vspace{0.2in}
	    \mbox{}
	    \includegraphics[width=.6\textwidth]{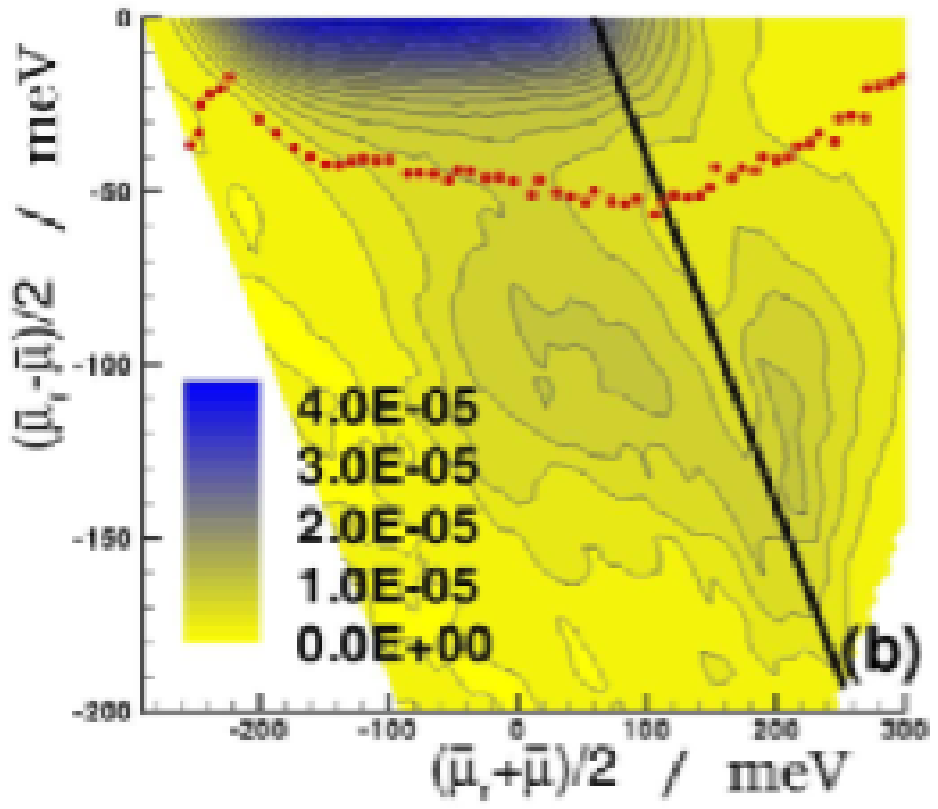}
	  \end{center}
	  \caption[First-order reversal curves (FORCs) and FORC
	    distribution for a discontinuous phase transition]{ (Color online.)
	    ({\bf a})
	    First-order reversal curves (FORCs) for a continuous phase
	    transition simulated at a high scan rate, $\Omega=0.01$\,meV/MCSS.
	    The black curve in the middle shows the equilibrium
	    isotherm. The minima of each FORC are also shown (black
	    filled circles).
	    ({\bf b})
	    FORC distribution generated from the FORCs shown in ({\bf a}).The
	    positions of the FORC minima are also shown (black filled circles).
	    The straight line corresponds to the FORC for which the minimum
	    lies closest to the critical coverage. 
	  }
	  \label{fig:highscan}
	\end{figure}

\begin{figure}
	  \vspace{0.4truecm}
	  \begin{center}
	    \includegraphics[width=.65\textwidth]{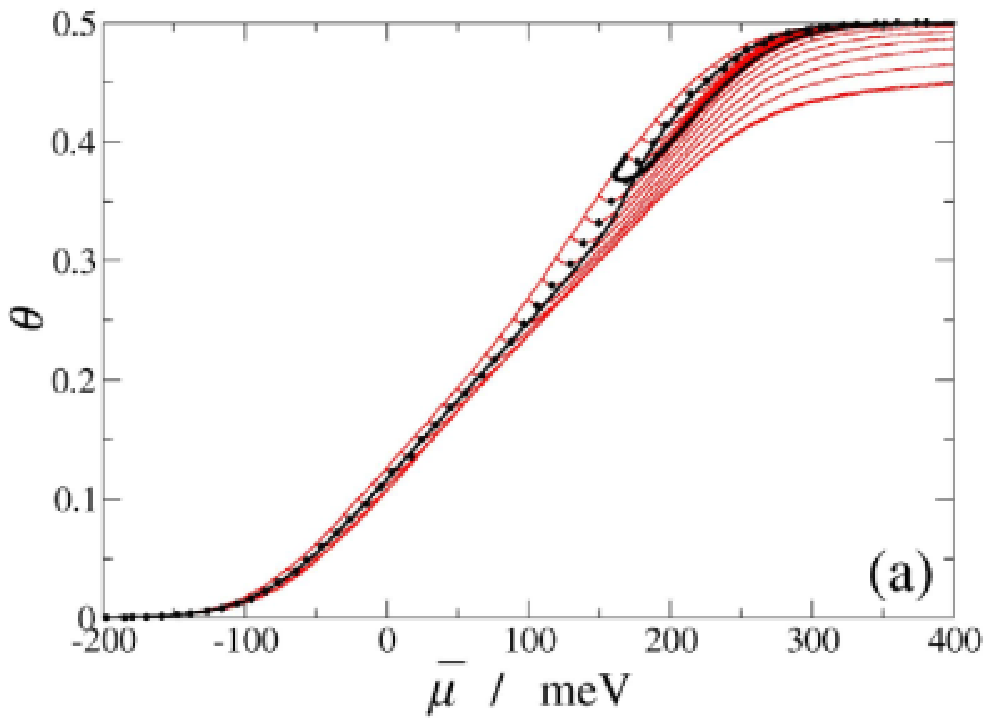}
	    \mbox{}
	    \vspace{0.2in}
	    \mbox{}
	    \includegraphics[width=.6\textwidth]{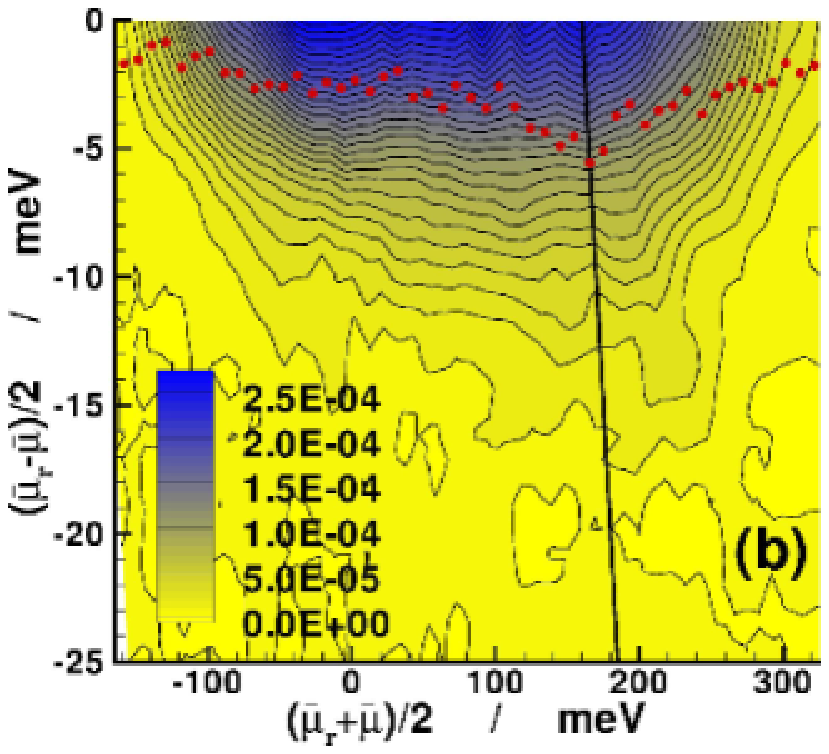}
	  \end{center}
	  \caption[First-order reversal curves (FORCs) and FORC
	    distribution for a discontinuous phase transition]{ (Color online.)
	    ({\bf a})
	    First-order reversal curves (FORCs) for a continuous phase
	    transition simulated with a large diffusion barrier
	    $\Delta_{\rm nn}=300$\, meV. The black curve in the middle
	    shows the equilibrium isotherm. The minima of each FORC
	    are also shown (black filled circles).
	    ({\bf b})
	    FORC distribution generated from the FORCs shown in ({\bf a}).The
	    positions of the FORC minima are also shown (black filled circles).
	    The straight line corresponds to the FORC for which the minimum
	    lies closest to the critical coverage.
	  }
	  \label{fig:lowdiff}
	\end{figure}

        The effect of jamming is more
	pronounced at higher scan rates, or with a higher diffusion
	barrier, where the rate of adsorption is much faster than the
	rate of diffusion. The family of FORCs and FORC diagram at a
	higher scan rate, $\Omega=0.01$\,meV/MCSS, are shown in
	Fig.~\ref{fig:highscan}, and the FORCs and FORC diagram with a
	larger diffusion barrier are shown in
	Fig.~\ref{fig:lowdiff}. In Fig.~\ref{fig:highscan}, two
	distinct groups of FORCs undergoing jammed and unjammed
	dynamics can be clearly seen. This is reflected in the FORC
	diagram as a splitting of the ``peninsula'' into two
	``islands'' of high $\rho$ values. A similar effect is
	seen in Fig.~\ref{fig:lowdiff}, since also there the rate of
	adsorption is much faster than the rate of diffusion (larger
	diffusion barrier). However, Fig.~\ref{fig:lowdiff}({\bf a}) shows a
	slight difference between the FORC minima and the equilibrium
	curve around the critical coverage. Notice also in
	Fig.~\ref{fig:highscan}({\bf a}) that even at a much higher
	scan rate than in Fig.~\ref{fig:fig2} (nearly two orders of
	magnitude), the FORC minima still follow the equilibrium curve
	very accurately. Thus, the FORC method should be useful to
	obtain the equilibrium adsorption isotherm quite accurately
        in experimental systems with slow equilibration rates.
	
	\section{Comparison and conclusions}
	\label{sec:conc}
	Two observations can be made by comparing the FORCs and FORC
	diagrams for systems with discontinuous and continuous phase
	transitions. First, the FORC minima in systems with a
	continuous phase transition correspond to the equilibrium
	behavior, while they do not for systems with a discontinuous
	phase transition. Thus, FORCs can be used to recover the
	equilibrium behavior for systems with continuous phase
	transitions that need a long time to equilibrate. This could
	be useful in experiments. Second, due to the instability that
	exists in systems with a discontinuous phase transition, the
	minima of the family of FORCs form a back-bending ``van der
	Waals loop'', and the corresponding FORC diagram contains
	negative regions which do not exist for systems with a
	continuous phase transition. Since experimental implementation
	of the FORC method should only require simple reprogramming of
	a potentiostat designed to carry out a standard CV experiment, we
	believe the method can be of significant use in obtaining
	additional dynamic as well as equilibrium information from
	such experiments for systems that exhibit electrochemical
	adsorption with related phase transitions.
	
	\section*{Acknowledgments}
	This research was supported by U.S. NSF Grant
	No. DMR-0240078, and by Florida State University through the School of 
	Computational Science, the Center for Materials Research and
	Technology, and the National High Magnetic Field Laboratory.
	
	\section*{Appendix}
	\appendix 
	In this appendix we present a mapping between lattice-gas models
	of adsorption and discrete spin models of magnetic systems, and
	then introduce the FORC method in the original magnetic language.
	
	The occupation variable in the lattice-gas model, $c_i \in \{0,1\}$
	is a binary variable, just like the magnetization variables:
	$s_i=M_s/m_s \in \{-1,1\} $ in the Classical Preisach Model (CPM). We
	therefore have the mappings $c_i=(s_i+1)/2$ and
	$(\bar{\mu}-\bar{\mu}_0) = 2 H$~\cite{Frank:06,pathria,Yang:52}.
	As a result, the FORC method can be applied to electrochemical
	adsorption, as well as to magnetic hysteresis.
	
\begin{figure}
	  \begin{center}
	    \includegraphics[width=.45\textwidth]{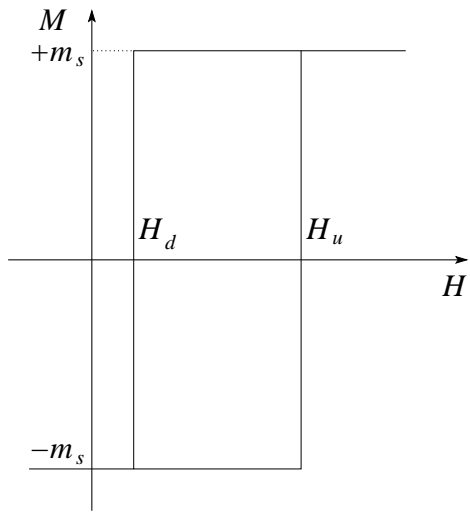}
	    \caption[Hysteresis loop for one hysteron]{
	      Schematic diagram of a hysteresis loop for a single hysteron.
	      $H_u$ and $H_d$
	      are the up and down reversal fields, respectively. After
	      Ref.~\cite{Mayergoyz:91}.
	    }
	    \label{fig:hysteron}
	  \end{center}
	\end{figure}

	The CPM is based on the
	idea that a material consists of a number of
	elementary interacting ``particles'' or ``domains,'' called
	hysterons. The hysterons are assumed to have rectangular hysteresis
	loops between two states that have the same magnetization
	values, $+m_s$ and $-m_s$, for all hysterons. A typical hysteresis
	loop for a hysteron is shown in Fig.~\ref{fig:hysteron}. $H_u$ and
	$H_d$ are the up and down switching magnetic fields
	respectively. It is  also assumed that the different hysterons
	have a distribution of reversal fields $\Phi(H_u,H_d)$. In the
	CPM, the total magnetization can be defined as~\cite{Mayergoyz:91}
	\begin{equation}
	  M(t)=\int \!\! \int_{H_u \ge H_d} \Phi(H_u,H_d)
	  [\hat{m}(H_u,H_d) H(t)] {\rm d}H_u {\rm d}H_d\; ,
	\end{equation}
	where the operator $\hat{m}(H_u,H_d)$ applied to $H(t)$ gives $+m_s$ if
	the particle is switched up and $-m_s$ if the particle is
	switched down. Note that $\hat{m}(H_u,H_d) H(t)$ depends on
	the field history of $H(t)$ and not only on the instantaneous
	value, which enables the CPM to model irreversible hysteresis
	behavior.
	For a detailed discussion
	of the Preisach Model, see Ref.~\cite{Mayergoyz:91}.
	
	In a typical FORC analysis of a magnetic system, the magnetization $M$
	is saturated in a positive applied magnetic field, and then the 
	applied magnetic field is decreased continuously to a reversal
	field $H_r$. The magnetic field is then increased back to saturation.
	A first-order reversal curve is the response of the magnetization
	to the increasing magnetic field ($H>H_r$). This is done for different
	values of $H_r$, and a set of curves, $M(H_r,H)$, is
	collected~\cite{kn:pike99,kn:pike03}. The FORC distribution is
	defined as
	\begin{equation}
	  \tilde{\rho}=- \frac{1}{2} \frac{\partial^2 M(H_r,H)}{\partial H_r\,
	    \partial H}\,,
	\end{equation} 
        where the tilde denotes a trivially different normalization
        from the one used here.
	Thus, using the mapping given above, one arrives at
	Eq.~(\ref{forc.definition}) as
	the definition of the FORC distribution in an electrochemical system.


\begin{thebibliography}{10}

\bibitem{Tansel:06}
T.~Tansel, O.~M. Magnussen, Phys. Rev. Lett. 96 (2006) 026101.

\bibitem{kn:mayergoyz86}
I.~D. Mayergoyz, IEEE Trans. Magn. MAG 22 (1986) 603.

\bibitem{kn:pike99}
C.~R. Pike, A.~P. Roberts, K.~L. Verosub, J. Appl. Phys.
  85 (1999) 6660.

\bibitem{kn:enachescu05}
C.~Enachescu, R.~Tanasa, A.~Stancu, F.~Varret, J.~Linares, E.~Codjovi,
  Phys. Rev. B 72 (2005) 054413.

\bibitem{kn:fecioru-morariu04}
M.~Fecioru-Morariu, D.~Ricinschi, P.~Postolache, C.~E. Ciomaga, A.~Stancu,
  J. Optoelectron. Adv. Mater. 6 (2004) 1059.

\bibitem{kn:robb05}
D.~T. Robb, M.~A. Novotny, P.~A. Rikvold, J. Appl. Phys. 97 (2005) 10E510.

\bibitem{AbouHamad:04}
I.~Abou~Hamad, P.~A. Rikvold, G.~Brown, Surf. Sci. 572 (2004) L355--L361.

\bibitem{MitchellSS:01}
S.~J. Mitchell, G.~Brown, P.~A. Rikvold, Surf. Sci. 471 (2001)
  125--142.

\bibitem{AbouHamad:03}
I.~Abou~Hamad, Th. Wandlowski, G. Brown, and P.~A. Rikvold, J. Electroanal. Chem. 554-555 (2003) 211--219.

\bibitem{AbouHamad:05}
I. Abou Hamad, S. J. Mitchell, Th. Wandlowski, and P. A. Rikvold, Electrochim. Acta 50 (2005) 5518--5525.

\bibitem{kn:pike03}
C.~R. Pike, Phys. Rev. B 68 (2003) 104424.

\bibitem{AbouHamad:06}
I.~Abou~Hamad, D.~T. Robb, P.~A. Rikvold, in: D.~P. Landau, S.~P. Lewis, H.-B. Sch\"{u}ttler (Eds.),
  Computer Simulation Studies in Condensed-Matter Physics XIX, Springer-Verlag,
  Berlin, in press.

\bibitem{Frank:05}
S.~Frank, D.~E. Roberts, P.~A. Rikvold, J. Chem. Phys. 122 (2005) 064705.

\bibitem{Frank:06}
S.~Frank, P.~A. Rikvold,  Surf. Sci., in press.

\bibitem{Mitchell:02}
S.~J. Mitchell, S.~Wang, P.~A. Rikvold, {Halide adsorption on single-crystal
  silver substrates: Dynamic simulations and \textit{ab initio} density
  functional theory}, Faraday Disc. 121 (2002) 53--69.

\bibitem{Brown:99}
G.~Brown, P.~A. Rikvold, S.~J. Mitchell, M.~A. Novotny, 
  in: A.~Wieckowski (Ed.), {Interfacial Electrochemistry: Theory, Experiment,
  and Application}, Marcel Dekker, New York, 1999, pp. 47--61.

\bibitem{Kang:89}
H.~C. Kang, W.~H. Weinberg, J. Chem. Phys. 90
  (1989) 2824--2830.

\bibitem{Buendia:04}
G.~M. Buend{\'\i}a, P.~A. Rikvold, K.~Park, M.~A. Novotny, J. Chem. Phys. 121 (2004) 4193--4202.

\bibitem{BrownJES:99}
G.~Brown, P.~A. Rikvold, M.~A. Novotny, A.~Wieckowski, J. Electrochem.
  Soc. 146 (1999) 1035.

\bibitem{Ocko:97}
{B. M. Ocko, J. X. Wang, and Th. Wandlowski}, Phys.
  Rev. Lett. 79 (1997) 1511--1514.

\bibitem{Wandlowski:01}
{Th. Wandlowski, J. X. Wang, and B. M. Ocko}, J. Electroanal. Chem. 500 (2001) 418--434.

\bibitem{Ramos:99}
R.~A. Ramos, P.~A. Rikvold, M.~A. Novotny, Phys. Rev. B 59 (1999) 9053--9069.

\bibitem{Savitzky:64}
A.~Savitzky, M.~J.~E. Golay, Anal. Chem. 36 (1964) 1627--1639.

\bibitem{Press:nr}
W.~H. Press, A.~Teukolsky, W.~T. Vetterling, B.~P. Flannery, Numerical Recipes
  in C: The Art of Scientific Computing, Cambridge University Press, Cambridge,
  1997.

\bibitem{Castellan}
G.~W. Castellan, Physical Chemistry, Addison-Wesley, Reading, MA, 1964.

\bibitem{Landau:00}
D.~P. Landau, K.~Binder, A Guide to Monte Carlo Simulations in Statistical
  Physics, Cambridge University Press, Cambridge, 2000.

\bibitem{Privman:93}
J.-S. Wand, P.~Nielaba, V.~Privman, Mod. Phys. Lett. B 7 (1993) 189.

\bibitem{pathria}
R.~K. Pathria, Statistical Mechanics, 2nd Edition, Butterworth Heineman,
  Oxford, UK, 1996.

\bibitem{Yang:52}
C.~N. Yang, T.~D. Lee, Phys. Rev. 87 (1952) 410.

\bibitem{Mayergoyz:91}
I.~D. Mayergoyz, Mathematical Models of Hysteresis, Springer-Verlag, New York,
  1991.

\end{thebibliography}
\end{document}